\documentclass[aps,prd,preprint]{revtex4-1}
\usepackage{epsfig}
\begin{document}

\title{Model Study of a Quark Star}
\author{Hua Li$^{1}$, Xin-Lian Luo$^{2}$, Yu Jiang$^{1}$, and Hong-Shi Zong$^{1,3}$}
\address{$^{1}$ Department of Physics, Nanjing University, Nanjing 210093, China}
\address{$^{2}$ Department of Astronomy, Nanjing University, Nanjing 210093, China}
\address{$^{3}$ Joint Center for Particle, Nuclear Physics and Cosmology, Nanjing 210093, China}

\begin{abstract}

In this paper we apply the equation of state (EOS) of QCD at finite chemical potential and
zero temperature proposed in H. S. Zong and W. M. Sun [Int. J. Mod. Phys. {\bf A23}, 3591 (2008)] to the study of properties of quark star. This EOS contains only one adjustable parameter $m_D$ which sets the scale of chiral symmetry breaking (in our calculation we have chosen two values of $m_D$: $m_D=244~MeV$ and $m_D=239~MeV$, which is fitted from the value of $f_\pi$ and determined by $e^+ e^-$ annihilation experiment, respectively).  
From this EOS a model of quark star is established by applying the Tolman-Oppenheimer-Volkoff equation under two conditions: with the $P(\mu=0)$ term and without the $P(\mu=0)$
term.  Our results show clearly that the $P(\mu=0)$ term is an important quantity in the study of quark star. A comparison between our model and other models of quark
star is made. In particular, we have compared our results with the most recent observational data measured using Shapiro delay reported in P.B. Demorest et al. [Nature (London) {\bf 467}, 1081 (2010)].

\bigskip

\bigskip


\end{abstract}
\maketitle

\section{introduction}
\label{S:intro}

Since Gell-Mann and Zweig \cite{Gell-Mann,Zweig} proposed the conception of quark, many authors have used this idea in the study of astrophysical problems \cite{Ivanenko,shapiro,Glendenning}. Just as the discovery of neutron led to the idea of neutron star, the conception of quark naturally stimulated many physicists and astrophysicists to suggest that quark star maybe exists as well \cite{Itoh,Iwamoto,Bodmer,Witten,Haensel,Alcock}. In the later decades the research about the nature of quark star was developed. As is well known, the equation of state (EOS) of matter plays a key role in the study of the structure of compact stars. However, at present it is very difficult to obtain the EOS of compact matter from the first principle of Quantum Chromodynamics (QCD) in all ranges of density and temperature. Therefore one has to adopt various QCD inspired models to study the properties of dense objects. For example: the MIT bag model \cite{Weisskopf,Weber,Soff,Paris}, the Nambu-Jona-Lasinio (NJL)model \cite{Rehberg,Greiner,Ruster,Menezes}, and the perturbative QCD model \cite{Freedman,Baluni,Fraga,Farhi}. Although all of these models can reflect some basic features of QCD, none of them can provide a complete and consistent framework to work in all ranges of density and temperature. The MIT bag model can provide the quark
confinement mechanism, the NJL model can describe chiral symmetry breaking of QCD, while the perturbative QCD model works well at high energy scale due to asymptotic freedom.
Present astrophysical observations and laboratory experimental data cannot
enable one to make a clear choice among these models. So it is worthwhile to develop other models with better QCD foundation to study compact matter. Recently, in Ref. \cite{zong2} the authors derived a model-independent formula for calculating the EOS of QCD at finite chemical potential and zero temperature with the aid of functional integral formalism and obtained an explicit analytic expression for the EOS by applying the rainbow-ladder approximation of the Dyson-Schwinger approach and the model quark propagator proposed in Ref. \cite{pagels}. In this paper we shall apply this EOS to the study of structure of quark stars by considering two situations: with the $P(\mu=0)$ term (the pressure at zero chemical potential and temperature) and without the $P(\mu=0)$ term.

This paper is organized as follows. In Sec. \ref{EOS}, we first give a brief introduction to the model-independent formula for the EOS of QCD at finite chemical potential and zero temperature derived in Ref. \cite{zong2} and give the explicit analytic expression of the EOS obtained using this formula with the rainbow-ladder approximation of the Dyson-Schwinger approach and the model quark propagator proposed in Ref. \cite{pagels}. From these results the properties of quark star, such as mass-radius and mass-energy-density relations are obtained by solving the Tolman-Oppenheimer-Volkoff (TOV) equation under two conditions: with the $P(\mu=0)$ term and without the $P(\mu=0)$ term. A comparison between our model and other models of
quark star is made. In Sec. \ref{discussion}, we offer a discussion of our study.

\section{Structure of Quark Matter}
\label{EOS}

(i) EOS of quark matter 

It is well-known that in equilibrium statistical field theory \cite{Kapusta}, once the partition function (or equivalently, EOS) of a system is obtained, then all the thermodynamical variables can be determined. The calculation of the EOS of QCD at finite chemical potential is a contemporary focus. However, due to dynamical chiral symmetry breaking and confinement, it is very difficult to get some valuable and reliable results about EOS of QCD from perturbation theory at finite chemical potential. So people have developed various nonperturbative methods for studying strong interaction physics at finite density and zero temperature \cite{Roberts,Schmidt,Roberts1,zong3,XY,Y}.
In Ref. \cite{zong2} the authors derived a model-independent formula for calculating the EOS of QCD at finite chemical potential and zero temperature with the aid of functional integral formalism. The expression of the EOS derived there reads
\begin{eqnarray}\label{EOSformula}
P(\mu)&=& P(\mu=0)+\int_{0}^{\mu}d \mu' \rho(\mu') \nonumber \\
&=& P(\mu=0)-N_c N_f \int_{0}^{\mu}d \mu' \int \frac{d^4 p}{(2\pi)^4}\mathrm{tr}\{G[\mu'](p)\gamma_4 \},
\end{eqnarray}
where $N_c$ and $N_f$ denote the number of quark colors and quark flavors, respectively; $G[\mu](p)$ and $\rho(\mu)=-N_c N_f \int \frac{d^4 p}{(2\pi)^4}\mathrm{tr}\{G[\mu'](p)\gamma_4 \}$
are the dressed quark propagator and the quark-number density at finite $\mu$, respectively. 
From Eq. (1) it can be seen that in order to obtain the EOS at finite chemical potential, a key issue is to know the behaviors of the dressed quark propagator at finite $\mu$. At present it is impossible to derive the exact quark propagator at finite $\mu$ from first principles of QCD. In Ref. \cite{Zong1} a new approximation method for obtaining the fermion propagator at finite $\mu$ was proposed by using the rainbow-ladder approximation of the Dyson-Schwinger approach and then the validity of this method was discussed in detail in Refs. \cite{Feng,He,sun,Feng2}.
According to Ref. \cite{Zong1}, in order to obtain the dressed quark propagator at finite $\mu$, one should specify the dressed quark propagator at $\mu=0$ in advance. In this work we choose the following model quark propagator proposed in Ref. \cite{pagels}:
\begin{equation}\label{PagelsProp}
 G(p)=\frac{1}{i\gamma \cdot p A(p^2)+B(p^2)} 
\end{equation}
with $A(p^2)=1,~B(p^2)=\frac{4 m_D^3}{p^2}$,
where $m_D$ is a parameter connected to a dynamically generated quark mass.
Then, according to the result in Ref. \cite{Zong1}, the dressed quark propagator at 
finite $\mu$ can be written as
\begin{equation}
G[\mu](p)=\frac{1}{i\gamma \cdot {\tilde p} A({\tilde p}^2)+B({\tilde p}^2)},
\end{equation} 
where ${\tilde p}=({\vec p}, p_4+i\mu)$.
Using the above $G[\mu](p)$, the authors in Ref. \cite{zong2} obtained an explicit analytic expression for the EOS by means of the formula (\ref{EOSformula}). 
The quark-number density obtained there is
\begin{eqnarray}
\label{NumDens} \rho(\mu)&=&\frac {N_c
N_f}{9\pi^2} \bigg( 2~\theta\big(\mu-
\frac{1}{2}4^{1/3}m_D \big)(1+\frac{3}{4}
16^{1/3}\frac{m_D^2}{\mu^2})^{3/2}(\mu^2-\frac{1}{4}16^{1/3}m_D^2)^{3/2}
 {} \nonumber\\
&&  {} +
\theta\big(\mu-4^{1/3}m_D\big)(\mu^2-16^{1/3}m_D^2)^{3/2}\bigg),
\end{eqnarray}
and the pressure $P(\mu)$ is
\begin{eqnarray}
\label{Pressure} P(\mu)&=&-\frac{N_c
N_f}{9\pi^2}16^{2/3}m_D^4~\theta\left(\mu-\frac{1}{2}4^{1/3}m_D\right)
(f(\frac{16^{1/3}m_D^2}{\mu^2})-f(4)){} \nonumber\\ &&
{}+\frac{N_c
N_f}{9\pi^2}16^{2/3}m_D^4~\theta\left(\mu-4^{1/3}m_D\right)g(\frac{\mu}{4^{1/3}}m_D)
+P(\mu=0),
\end{eqnarray}
where
\begin{eqnarray}
\label{f}
f(\mu)&=&\frac{1}{64}\bigg((-3-\frac{8}{\mu^2}-\frac{10}{\mu})\sqrt{16+8\mu-3\mu^2}
{} \nonumber\\ && {}
+12\sqrt{3}\arcsin\frac{4-3\mu}{8}+12\ln\frac{4+\mu+\sqrt{16+8\mu-3\mu^2}}{\mu}\bigg);
\end{eqnarray}
\begin{equation}
\label{g}
g(\mu)=\frac{1}{8}\bigg(\mu\sqrt{\mu^2-1}(2\mu^2-5)+3\ln(\mu+\sqrt{\mu^2-1})\bigg);
\end{equation}
and $ P(\mu=0)$ (the pressure at zero chemical potential) is calculated from the Cornwall- Jackiw-Tomboulis (CJT) effective
action \cite{Cornwall, Stam}
\begin{eqnarray}
\label{p} 
P(\mu=0) &=& N_c N_f \int \frac{d^4 p}{(2\pi)^4}\bigg\{ \rm{tr} \ln \big[G_0(p)G^{-1}(p)\big]-\frac{1}{2}\rm{tr}\big[1-G_0^{-1}(p)G(p)\big]\bigg\} \nonumber \\
&=& 2 N_c N_f \int \frac{d^4 p}{(2\pi)^4} \bigg\{ \ln\big [\frac{A^2(p^2)p^2+B^2(p^2)}{p^2}\big] -\frac{p^2 A(p^2)[A(p^2)-1]+B^2(p^2)}{p^2 A^2(p^2)+B^2(p^2)}\bigg \},
\end{eqnarray}
where $G_0(p)$ is the bare quark propagator.
Substituting the model quark propagator (\ref{PagelsProp}) into Eq. (\ref{p}), one obtains
\begin{equation}
P(\mu=0)=\frac{2N_c N_f}{3^{3/2}16^{1/3}\pi}m_D^4.
\end{equation}
Here, we want to stress that $m_D$ is connected to the dynamically generated quark mass.
When one discusses low energy QCD at finite $\mu$, it is evident that the dynamically generated quark mass should be $\mu$ dependent. So, in principle, $m_D$ should depend on $\mu$. However, in this work, according to the approximation adopted in Ref. \cite{zong2}, $m_D$ is taken to be $\mu$ independent (In fact, the value of $m_D$ can be determined experimentally, for example, from the value of $\pi$ decay constant $f_\pi$ or determined by $e^+ e^-$ annihilation experiment).

Before we turn to the study of quark star by means of EOS (\ref{Pressure}), it is necessary to have a look at EOS (\ref{Pressure}) itself. 

First let us see the large $\mu$ behavior of $P(\mu)$. For $\mu > 4^{1/3} m_D$, the two step functions in EOS (\ref{Pressure}) equal 1. 
Using the explicit expressions of $f(\mu)$ and $g(\mu)$, one can expand the pressure in EOS (\ref{Pressure}) around $\mu=\infty$ as follows:
\begin{eqnarray}\label{nonpEOSasymp}
P(\mu) &=& N_c N_f \bigg[ \frac{1}{12\pi^2} \mu^4+\frac{m_D^4}{16^{1/3}\pi^2} +\frac{m_D^6}{12\pi^2}\frac{1}{\mu^2}+\cdots  \bigg] \nonumber \\
&=& \frac{N_c N_f}{12\pi^2}\mu^4 \bigg[ 1+\frac{12}{16^{1/3}} \bigg(\frac{m_D}{\mu}\bigg)^4+\bigg(\frac{m_D}{\mu}\bigg)^6+\cdots \bigg].
\end{eqnarray}
This equation clearly shows that, as $\mu \rightarrow \infty$, $P(\mu)$ tends to the pressure  of a massless free quark gas $P_0 (\mu)=\frac{N_c N_f}{12\pi^2}\mu^4$, and the deviation from the free quark gas result can be expanded in powers of the ratio $m_D/\mu$. Here it is  interesting to compare our EOS (\ref{Pressure}) with the cold, perturbative EOS proposed in Ref. \cite{Fraga}. The pressure density to second order in $\alpha_s$ in the $\overline{\rm{MS}}$ scheme obtained in Ref. \cite{Fraga} is quoted as
follows:
\begin{equation}\label{FPS}
P_{FPS}(\mu)=\frac{N_f \mu^4}{4\pi^2}\bigg \{1-2\bigg(\frac{\alpha_s}{\pi}\bigg) -
\bigg[G+N_f \ln \frac{\alpha_s}{\pi} +\bigg(11-\frac{2}{3}N_f \bigg)\ln \frac{\bar{\Lambda}}{\mu} \bigg]  \bigg(\frac{\alpha_s}{\pi}\bigg)^2 \bigg\},
\end{equation}
where $G=G_0 -0.536N_f+N_f \ln N_f,G_0=10.374\pm 0.13$ and $\bar{\Lambda}$ is the renormalization subtraction point. The scale dependence of the strong coupling constant 
$\alpha_s({\bar \Lambda})$ is taken as
\begin{equation}
\alpha_s({\bar \Lambda})=\frac{4\pi}{\beta_0 u}\bigg[ 1-\frac{2\beta_1}{\beta_0^2}\frac{\ln u}{u} +\frac{4\beta_1^2}{\beta_0^4 u^2}\bigg( \bigg(\ln u-\frac{1}{2}\bigg)^2 +
\frac{\beta_2 \beta_0}{8 \beta_1^2} -\frac{5}{4}\bigg)\bigg],
\end{equation}
where $u=\ln(\bar{\Lambda}^2/\Lambda^2_{\overline{\rm{MS}}}),~\beta_0=11-2N_f/3, 
~\beta_1=51-19N_f/3$ and $\beta_2=2857-5033 N_f/9+325 N_f^2/27$. 
For $N_f=3,~\Lambda_{\overline{\rm{MS}}}=365 ~\rm{MeV}$. 
The only freedom in the model of Ref. \cite{Fraga} is the choice of the ratio ${\bar \Lambda}/\mu$, which is taken to be 2 in that reference. The perturbative EOS (\ref{FPS}) is applicable only in the chirally symmetric phase, when the chemical potential $\mu$ is larger than $\mu_\chi$, the chiral phase transition point. From Eq. (\ref{FPS}) it can be also seen that as $\mu \rightarrow \infty$, the perturbative EOS also tends to the free gas result, and the correction terms are from perturbative contributions. By comparing Eqs. (\ref{Pressure}) and (\ref{FPS}), it can be seen that for moderate values of $\mu$ there is a quantitative difference between our EOS (\ref{Pressure}) and the perturbative EOS (\ref{FPS}): for our EOS the correction terms are positive and the pressure is larger than the free quark gas result, whereas for the perturbative EOS the correction terms are negative and the pressure is lower than the free quark gas result. Here it also should be noted that in our model $m_d$ sets the scale for dynamical chiral symmetry breaking, therefore the correction terms in our EOS arise from nonperturbative effects. 

Second, let us see the small $\mu$ behavior of our EOS (\ref{Pressure}). 
From Eq. (2) it can be seen that the obtained quark-number density distribution differs significantly from the Fermi distribution of a free quark gas. Physically this is a consequence of dynamical chiral symmetry breaking in the low energy region. We note that when $\mu$ is smaller than a critical value $\mu_0=\frac{1}{2}4^{1/3}m_D$, the quark-number density vanishes identically, and the pressure $P(\mu)$ equals $P(\mu=0)$ (i.e., the pressure of the vacuum). Namely, $\mu=\mu_0$ is a singularity which separates two regions with different quark-number densities. This result agrees qualitatively with the general 
conclusion of Ref. \cite{Jackson}. In that reference, based on a universal argument, it is pointed out that the existence of some singularity at the point $\mu=\mu_0$ and $T=0$ is a robust and model-independent prediction. It should be noted that any EOS of QCD at finite chemical potential and zero temperature with good QCD foundation should at least satisfy this requirement and our EOS does satisfy it.

From the above analysis it can be seen that in both the large $\mu$ limit and small $\mu$
limit our EOS has the correct behaviors required by QCD, which is quite different from other  EOSs, such as the perturbative EOS (\ref{FPS}). In addition, our EOS (3) has another merit that it has a simple analytic form and contains only one adjustable parameter $m_D$. So it can be easily applied in practical calculation of quark stars. Therefore, we expect that our EOS (3) can provide a possible new approach for the study of quark stars.

Just as was shown in Ref. \cite{Li1}, the $P(\mu=0)$ term in the EOS is an important quantity, since it enters the energy density, which is relevant for integrating the TOV equations. In order to see explicitly the influence of the $P(\mu=0)$ term to the properties of quark star, in the calculations that follow, we shall distinguish two situations: with the $P(\mu=0)$ term and without the $P(\mu=0)$ term. A discussion on the physical meaning of this term will be given in the final part of this paper.

\begin{figure}
\vspace{5mm}\centerline{\epsfig{file=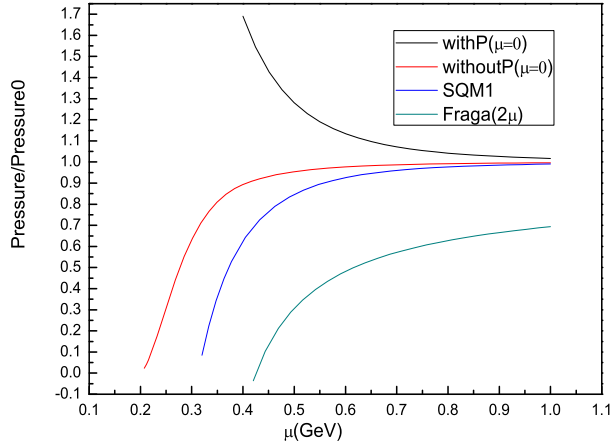,width=100mm,height=72.5mm} }
\vspace{-10mm}
\caption{The pressure as a function of chemical potential, relative to the free
quark gas pressure $P_{free}=N_cN_f\mu^4/(12\pi^2)$. In the SQM1,
$B=(164.34MeV)^4$
 \label{fig:P}}
\end{figure}

\begin{figure}
\vspace{5mm}\centerline{\epsfig{file=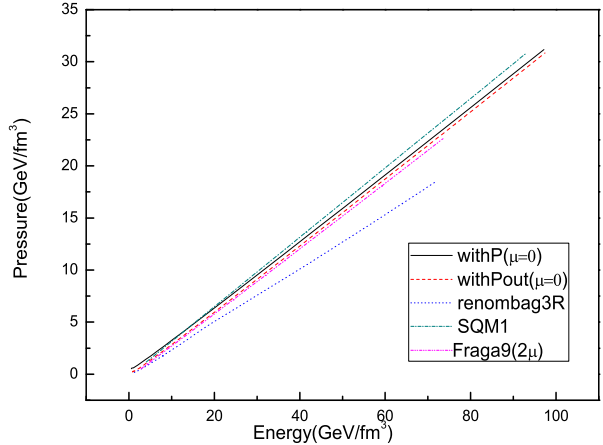,width=100mm,height=72.5mm} }
\vspace{-10mm}
\caption{Quark pressure-energy relation.
 \label{fig:Pressure-Energy}}
\end{figure}

Now let us turn to using EOS (3) to study the properties of quark star. In order to see the behavior of the pressure with the variation of $\mu$, we plot the $P-\mu$ curve in Fig. \ref{fig:P}. Here, as in Ref. [25], we take $m_D=244 ~MeV$ in our calculation. 
From Fig. \ref{fig:P} it can be seen that the pressure decreases
with $\mu$ increasing under condition one (with the $P(\mu=0)$ term), whereas it increases with $\mu$ increasing under condition two (without the $P(\mu=0)$ term). The results of
Fraga et. al. \cite{Fraga} and SQM1 \cite{Baron} are also shown in this figure.
In Fig. \ref{fig:Pressure-Energy} we make a comparison of the results in the present
paper with those in our previous paper \cite{Li1}, Fraga et. al. \cite{Fraga} and the SQM1 models \cite{Baron}. It can be seen that the SQM EOS is the stiffest and the renombag3R model EOS \cite{Li1} is the softest. This can be easily understood since the
interaction between particles will soften the EOS of the system.

(ii) Property of quark star 

Using the above results, one can obtain the structure of quark star
by integrating the TOV equation:
\begin{equation}
\label{TOV1} \frac{dP(r)}{dr}=-\frac{G~(\varepsilon+P)(M+4 \pi
r^3P)}{r(r-2GM)},
\end{equation}
\begin{equation}
\label{TOV2} \frac{dM(r)}{dr}=4\pi r^2\varepsilon .
\end{equation}
The calculated mass-radius relations and mass-energy density relations are shown in Figs. 3-6, respectively. 
Here, in order to see the sensitivity of the mass-radius relation on the parameter $m_D$ in our EOS (3), in the following calculation we will choose two values of $m_D$: $m_D=244~MeV$
and $m_D=239~MeV$, which is fitted from the value of $f_\pi$ \cite{pagels} and determined by $e^+ e^-$ annihilation experiment \cite{Sanda}, respectively.
In order to see clearly the calculated results in Figs. 3-6, we list them in Table I. 

\begin{table}
\caption{Model parameters and calculational results} {\label{tab:a}}
\begin{tabular}{|l|l|c|c|l|l|c|c|}
\hline 
\multicolumn{4}{|c|}{Without $P(\mu=0)$ term} &
\multicolumn{4}{c|}{With $P(\mu=0)$ term} \\
\hline
$m_d (MeV)$ & $M_\odot$ & R(km) & $\rho(g/cm^3)$ & $m_d (MeV)$ & $M_\odot$ & R(km) &
$\rho(g/cm^3)$ \\
\hline
 $244$& $2.3$ & $24$ & $6.76 \times 10^{14}$ & $244$ &$ 0.80$ & $8.0$ &
$10^{16}$
 \\
\hline
 $239$& $1.8$ & $17$ & $5.22 \times 10^{14}$ & $239$ & $0.87$ & $8.4$ &
$9.33 \times 10^{15}$
\\
\hline
\end{tabular}
\end{table}

From Table I it can be seen that, for the case without the $P(\mu=0)$ term, for $m_D=244~MeV$, the maximal mass $\sim 2.3~M_\odot$, the radius $\sim 24~km$ and the mass density $\sim 6.76 \times 10^{14}~g/cm^3 $, which is similar to the results calculated by means of the EOS proposed in Ref. \cite{zong3} without the $P(\mu=0)$ term. The curve of the mass-radius relation is the same as that of a neutron star. For $m_D=239~MeV$, the maximal mass $\sim 1.8~M_\odot$, the radius $\sim 17~km$ and the mass density $\sim 5.22 \times 10^{14}~g/cm^3 $.
Here we note that our results for $m_D=239~MeV$ are very close to the most recent observational data reported in Ref. \cite{Ransom} (in that reference the mass of the pulsar J1614-2230 is calculated to be $(1.97\pm 0.04)M_\odot$ and its radius is between 11 and 15 km). As was pointed out in that reference, this result rules out almost all currently proposed hyperon or boson condensate EOS, whereas quark matter can support a star of such a mass only if the quarks are strongly interacting and are therefore not "free" quarks. Our results support this observational data. For the case with the $P(\mu=0)$ term, for $m_D=244~MeV$, the maximal mass $\sim 0.8~M_\odot$, the radius $\sim 8~km$ and the mass density $\sim 1.06 \times 10^{16}~g/cm^3$. For $m_D=239~MeV$, the maximal mass $\sim 0.87~M_\odot$, the radius $\sim 8.4~km$ and the mass density $\sim 9.33 \times 10^{15}~g/cm^3 $. For these two values of $m_D$, the curve of the mass-radius relation is the same as that of a typical quark star. From the above results it can be seen that for the case without the $P(\mu=0)$ term, the structure of the star is sensitive to the value of $m_D$, whereas for the case with the $P(\mu=0)$ term, the structure of the star is insensitive to the value of $m_D$. 

From Figs. 5 and 6 it can be seen that the mass density obtained without the $P(\mu=0)$ term is less than the corresponding one obtained with the $P(\mu=0)$ term. The mass density obtained with the $P(\mu=0)$ term is larger than the corresponding ones in all other models ($\rho\sim 10^{16}\frac{g}{cm^3}$). This is because the EOS for this case is soft and leads to a more compact star. Here it is interesting to compare the study in the present paper with that in a previous work by some of the same authors \cite{Li1}. In Ref. \cite{Li1}, the $P(\mu=0)$ term is identified with $-B$, where $B$ is the vacuum bag constant, and $B$ is taken as a phenomenological input, whereas in the present paper the $P(\mu=0)$ term is determined self-consistently in the rainbow-ladder approximation of the Dyson-Schwinger approach.  

\begin{figure}
\vspace{5mm}\centerline{
\epsfig{file=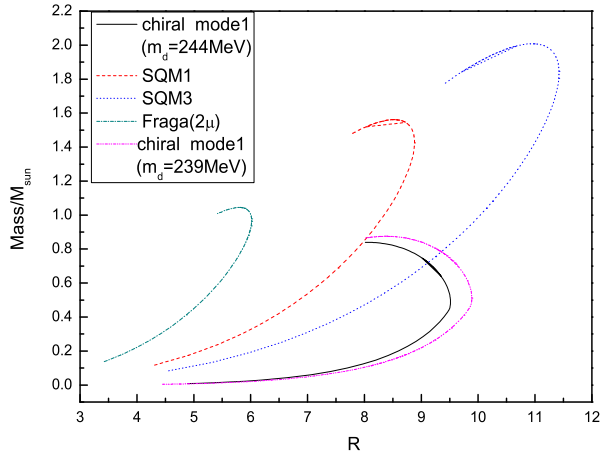,width=100mm,height=72.5mm} }
\vspace{-10mm}
\caption{Mass as a function of radius.
\label{fig:M-R1}}
\end{figure}
\begin{figure}
\vspace{5mm}\centerline{\epsfig{file=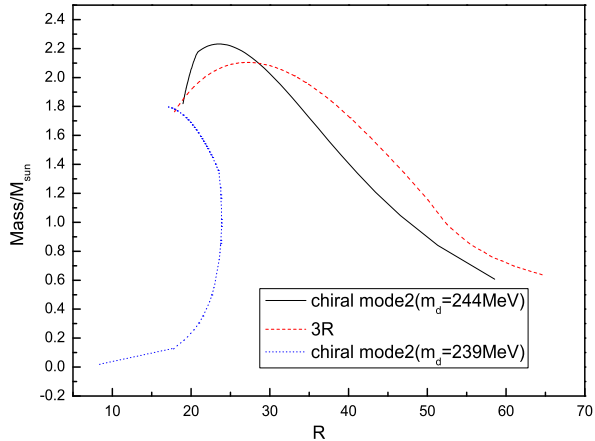,width=100mm,height=72.5mm} }
\caption{Mass-radius relation. The 3R curve is calculated from the EOS proposed in Ref. \cite{zong3} without the $P(\mu=0)$ term.
\label{fig:M-R2}}
\end{figure}
\begin{figure}
\vspace{5mm}\centerline{\epsfig{file=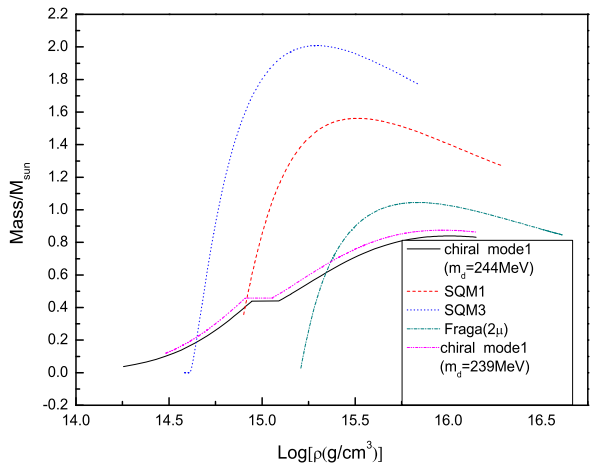,width=100mm,height=72.5mm} }
\vspace{-10mm}
\caption{Mass-energy density relation.
\label{fig:M-rho1}}
\end{figure}
\begin{figure}
\vspace{5mm}\centerline{\epsfig{file=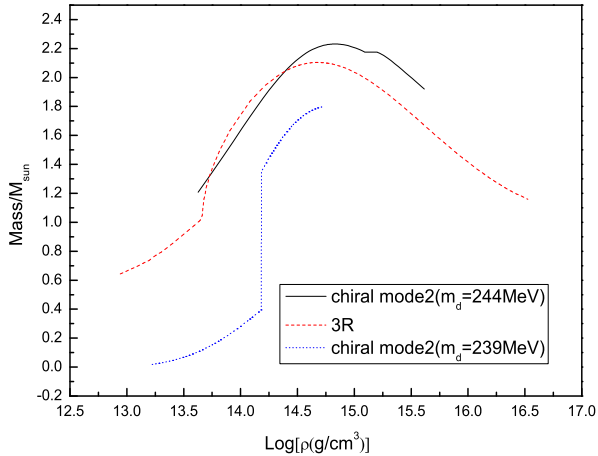,width=100mm,height=72.5mm} }
\vspace{-10mm}
\caption{Mass-energy density relation. The 3R curve is calculated from the EOS proposed in Ref. \cite{zong3} without the $P(\mu=0)$ term.
 \label{fig:M-rho2}}
\end{figure}

\section{Discussion}
\label{discussion}

As we have mentioned above, in the present paper we consider two conditions: with the $P(\mu=0)$ term and without the $P(\mu=0)$ term (the $P(\mu=0)$ term is calculated using the CJT effective action). This term is not the same
as the bag constant $B$ in the MIT bag model, because it is positive while $B$ is
negative! Furthermore, in the calculation of structure of stars this
term leads to negative energy density at small chemical
potential, which is hard to understand from a physical point of view. This is one of our motivations for adopting the EOS proposed in Ref. \cite{zong2} to study the properties of quark star. In addition, in spite of the progress made in the
study of quark stars, there are still some authors who suspect the
existence of quark star. In Refs. \cite{Serot,Baym,Kisslinger},
the authors think that there is no quark matter core in the
neutron star. \"Ozel \cite{Ozel} argued from their observational
data that quark star cannot exist, because in term of their
observational results the mass $\geq 2.1~\pm 0.28 M_\odot$ and the radius $R~\geq 13.8~\pm 1.8 Km$. Of course, other authors do not consent to her arguments \cite{Drago}, and
the discovery of meta-millisecond pulsar is in favor of quark star \cite{Baron}. Here we also note that in a recent paper \cite{Ransom}, the authors argue that their observational data support the existence of quark star with strong interactions. Therefore, the present astronomical observations do not rule out the existence of quark stars. 

To summarize, in this paper we adopt the EOS of QCD at finite chemical potential and zero temperature proposed in Ref. \cite{zong2} to study the properties of quark star. A comparison between our result and those in the literatures is made. Applying this EOS, one obtains the structure of quark star. The results obtained under the two
conditions (with the $P(\mu=0)$ term and without the $P(\mu=0)$ term) are very different and the reason for this is analyzed. This shows that the $P(\mu=0)$ term plays an important role in the study of properties of quark star.
\begin{acknowledgments}

This work is supported in part by the National Natural Science
Foundation of China under Grant Nos 10775069, 10935001 and 11075075 and the Research
Fund for the Doctoral Program of Higher Education under Grant No
200802840009.

\end{acknowledgments}

\end{document}